\documentclass[aps,prb,twocolumn,showpacs]{revtex4}

\usepackage{graphicx}
\usepackage{dcolumn}
\usepackage{amsmath}
\usepackage{amssymb}
\usepackage{wasysym}
\usepackage{color}
\begin{document}

\title{The number of holes contained within the Fermi surface volume in underdoped high temperature superconductors}

\author{N.~Harrison%\dots%$^1$, S.~E.~Sebastian$^2$
}

\affiliation{%$^1$
Mail~Stop~E536,~Los~Alamos~National Labs.,Los~Alamos,~NM~ 87545%\\
%$^2$Cavendish Laboratory, Cambridge University, JJ Thomson Avenue, Cambridge CB3~OHE, U.K
}
\date{\today}

\begin{abstract}
We bring resolution to the longstanding problem relating Fermi surface reconstruction to the number of holes contained within the Fermi surface volume in underdoped high $T_{\rm c}$ superconductors. On considering uniaxial and biaxial charge-density wave order, we show that there exists a relationship between the ordering wave vector, the hole doping and the cross-sectional area of the reconstructed Fermi surface whose precise form depends on the volume of the starting Fermi surface. We consider a `large' starting Fermi surface comprising $1+p$ hole carriers, as predicted by band structure calculations, and a `small' starting Fermi surface comprising $p$ hole carriers, as proposed in models in which the Coulomb repulsion remains the dominant energy. Using the reconstructed Fermi surface cross-sectional area obtained in quantum oscillation experiments in YBa$_2$Cu$_3$O$_{6+x}$ and HgBa$_2$CuO$_{4+x}$ and the established methods for estimating the chemical hole doping, we find the ordering vectors obtained from x-ray scattering measurements to show a close correspondence with those expected for the small starting Fermi surface. We therefore show the Coulomb repulsion to remain largely unscreened throughout the entire underdoped regime where the pseudogap exists and further show that the quantum oscillation frequency and charge-density wave vectors provide accurate estimates for the number of holes contributing to the Fermi surface volume.
\end{abstract}
\pacs{71.45.Lr, 74.72.-h, 74.72.Gh, 74.72.Kf}
\maketitle

\section{Introduction}
\begin{figure}[ht!!!!!!!!!!] 
\centering 
\includegraphics*[width=.43\textwidth]{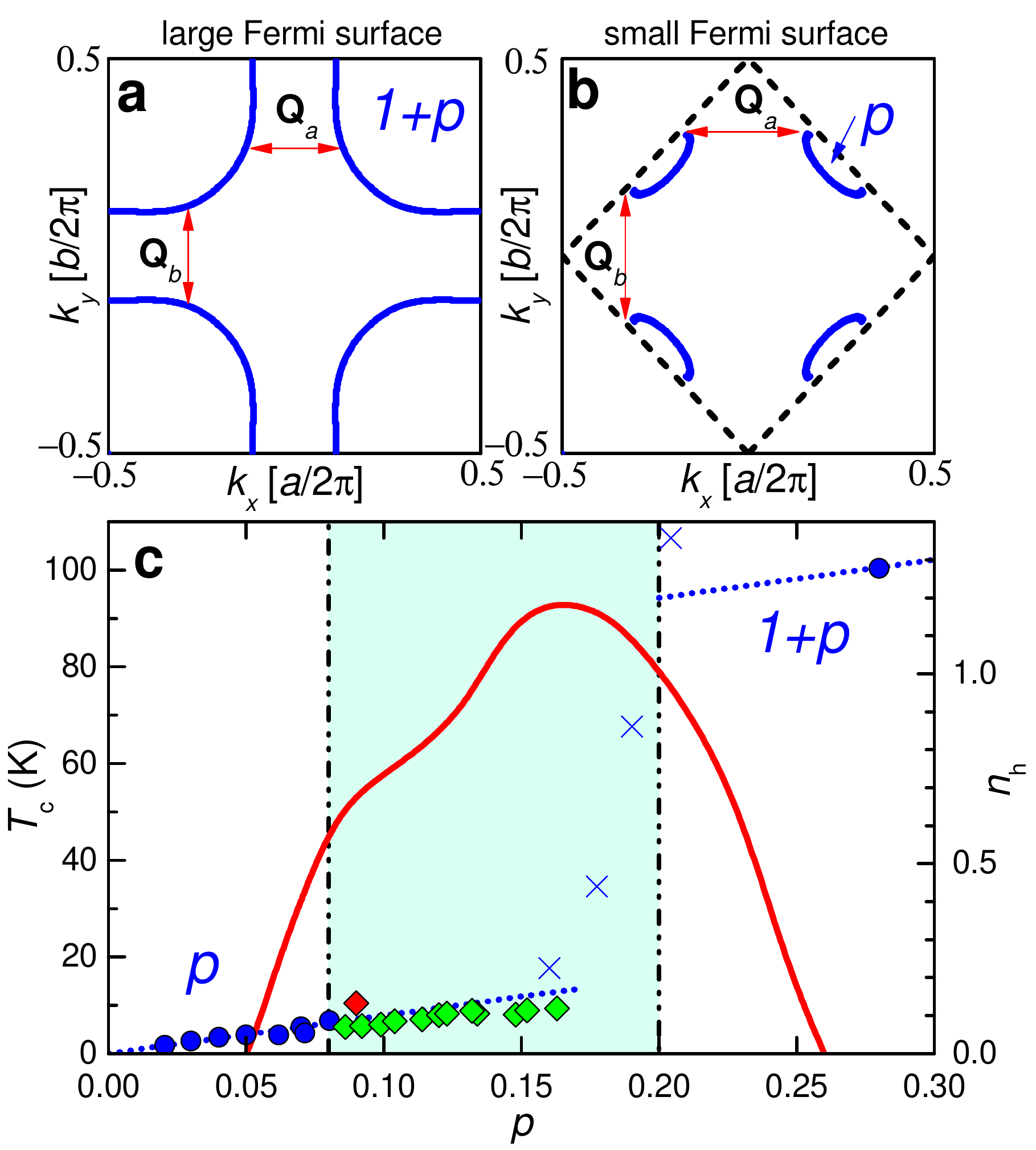}
\caption{({\bf a}), A schematic of the unreconstructed large cuprate hole Fermi surface\cite{andersen1} (neglecting bilayer coupling) that contains $1+p$ holes per unit cell. ${\bf Q}_a$ and ${\bf Q}_b$ illustrate notional charge-density wave ordering vectors. ({\bf b}), A schematic of a small Fermi surface, in which four small hole pockets bounded by the antiferromagnetic Brillouin zone boundary (dotted line) together contain $p$ holes per unit cell. ({\bf c}, left-hand axis) Notional doping-dependence of $T_{\rm c}$ (red curve).\cite{liang1} The Fermi surface volume of the intermediate doping range (shaded in cyan) has not previously been ascertained. ({\bf c}, right-hand axis) Experimental estimate of the number of holes, $n_{\rm h}=2A_{p,1+p}/A_{\rm UBZ}$, contributing to the Fermi surface volume. Green and red diamonds indicate $n_{\rm h}$ estimated from the quantum oscillation frequency and charge-density wave vectors using Equation~(\ref{inverse}) in YBa$_2$Cu$_3$O$_{6+x}$ and HgBa$_2$CuO$_{4+x}$, respectively. For $p<$~0.1, we assume $\delta_a=\delta_b$. Blue circles indicate the results of earlier Hall effect, angle-dependent magnetoresistance and quantum oscillation measurements,\cite{vignolle1,hussey1,segawa1} while crosses indicate the recent Hall results of Badoux {\it et al.}.\cite{badoux1} The dotted line represents $n_{\rm h}=p$ (for $p<0.2$) and $n_{\rm h}=1+p$ (for $p\geq0.2$) expected for a small and large Fermi surface, respectively.
}
\label{startingFS}
\end{figure}

The pseudogap is central to our understanding of high temperature superconductivity in the cuprates,\cite{imada1,dagotto1,timusk1} yet the number of hole carriers contained within the Fermi surface volume has remained challenging to ascertain experimentally. At issue is the degree to which Coulomb interactions cause the pseudogap to depart from a conventional metallic state. In the case of a conventional metal, a `large' Fermi surface volume consistent with band structure calculations is expected to result when Coulomb interactions between carriers are screened. In the cuprates, this large Fermi surface comprises $n_{\rm h}=1+p$ hole carriers (see Fig.~\ref{startingFS}a),\cite{andersen1} where, by convention, $p$ is the hole doping defined relative to the half filled band. In the case of a more unconventional metal, by contrast, the on-site Coulomb repulsion is largely unscreened causing it to dominate over low energy excitations. In this case, antiferromagnetic correlations are expected to remove one hole per CuO$_2$ plane per unit cell, leading to a `small' Fermi surface comprising $n_{\rm h}=p$ hole carriers.\cite{chakravarty1,lee1,rice1,qi1,chowdhury1} One of the possible outcomes is a small Fermi surface consisting of four hole pockets located at the antiferromagnetic Brillouin zone boundary (see Fig.~\ref{startingFS}b).

The large and small Fermi surface volumes have both been reported in the experiments, but at opposite ends of the doping phase diagram and outside of the pseudogap regime (see Fig.~\ref{startingFS}c). Deep in the overdoped regime at hole dopings $p\gtrsim$~0.20, Hall effect,\cite{badoux1} magnetic quantum oscillation\cite{vignolle1} and angle-dependent magnetoresistance oscillation\cite{hussey1} measurements are all found to be consistent with the large Fermi surface. Deep in the underdoped regime at very low hole dopings, $p\lesssim$~0.08, meanwhile, Hall effect measurements\cite{segawa1} are found to be consistent with the small Fermi surface. The presence of static staggered moments at these same very low dopings implies that the small Fermi surface there is the product of antiferromagnetism. 

The pseudogap regime, for which the total volume of the Fermi surface has remained undetermined,\cite{balakirev1} spans a broad intermediate range of hole dopings 0.08~$\lesssim p\lesssim$~0.20 (see Fig.~\ref{startingFS}c). The low temperature Hall effect has been found to be negative over much of this range in the highest quality samples,\cite{leboeuf1,doiron2} indicating it no longer to provide a direct measure of the number of holes contributing to the Fermi surface. X-ray scattering and nuclear magnetic resonance experiments have further revealed the presence of charge-density waves over most of this range rather than antiferromagnetism,\cite{wu1,ghiringhelli1,chang1,blackburn1,blanco1,tabis1,gerber1,chang2} with a possible broken rotational symmetry.

A biaxial charge-density wave order with two concurrent orthogonal wave vectors, ${\bf Q}_a=(\delta_a,0)\frac{2\pi}{a}$ and ${\bf Q}_b=(0,\delta_b)\frac{2\pi}{b}$, has been shown account for a large body of experimental data relating to the reconstructed Fermi surface within the pseudogap regime.\cite{harrison1,harrison2,sebastian1,sebastian2,maharaj1,allais1,harrison3,robinson1,briffa1} This data includes the small Fermi surface cross-sectional area found in quantum oscillation experiments,~\cite{doiron1,yelland1,bangura1,barisic1} the negative value of the Hall coefficient at high magnetic fields\cite{leboeuf1,doiron2} and the small value of the electronic heat capacity at high magnetic fields.\cite{riggs1,marcenat1} It has continued to remain unclear, however, as to whether it is a large starting Fermi surface (like that in Fig.~\ref{startingFS}a) or a small starting Fermi surface (similar to that in Fig.~\ref{startingFS}b) that becomes reconstructed by the charge-density wave.\cite{wu1,ghiringhelli1,chang1,harrison1,harrison2,sebastian1,sebastian2,maharaj1,allais1,harrison3,robinson1,briffa1,wise1,comin1}

Here we show that the observed reconstructed Fermi surface consisting mostly of a single electron pocket per CuO$_2$ plane\cite{harrison3,chan1,hsu1} and the measured values of the charge-density wave vectors\cite{ghiringhelli1,chang1,blackburn1,blanco1,tabis1} together point conclusively to a small starting Fermi surface (see Fig.~\ref{startingFS}c). 
We show using geometry that there exists a simple expression for the dependence of the length of ordering vector $\delta_{a,b}$ on hole doping $p$ and the momentum-space cross-sectional area of the reconstructed pocket(s) $A_{\rm e}$. Here, $\delta_a$ and $\delta_b$ are defined relative to the lengths of the unreconstructed Brillouin zone reciprocal lattice vectors ${\bf K}_a=(\frac{2\pi}{a},0)$ and ${\bf K}_b=(0,\frac{2\pi}{b})$. The functional form of $\delta_{a,b}$ on $p$ and $A_{\rm e}$ is sufficiently different for a large and small starting Fermi surface, that it can unambiguously distinguish between these scenarios. 
We therefore find $A_{\rm e}$ and $\delta_{a,b}$ to provide a reliable experimental means for estimating $n_{\rm h}$ over the majority of the pseudogap regime (see green and red diamonds in Fig.~\ref{startingFS}c).

%We find the ordering vector lengths $\delta_{a,b}$ reported in x-ray scattering\cite{ghiringhelli1,chang1,blackburn1,blanco1,tabis1} to lie closer to $\delta^{\rm small}$, as expected for a small Fermi surface. 
%

\section{Derivation}
The geometrical origin of the dependence of $\delta_{a,b}$ on $p$ and $A_{\rm e}$ can be visualized by considering an idealized form for the unreconstructed Fermi surface, such as that expected to apply in YBa$_2$Cu$_3$O$_{6+x}$ (shown in Fig.~\ref{startingFS}) when bilayer coupling and higher order hopping terms are neglected. Below we show that the derived expression for $\delta_{a,b}$ as a function of $p$ and $A_{\rm e}$ remains robust against an increase in strength of the charge-density wave order from weak to strong coupling. We also show it to remain robust against the introduction of bilayer hopping terms and changes in Fermi surface shape.

\subsection{Open Fermi surface and uniaxial order}
\begin{figure}
\centering 
\includegraphics*[width=.45\textwidth]{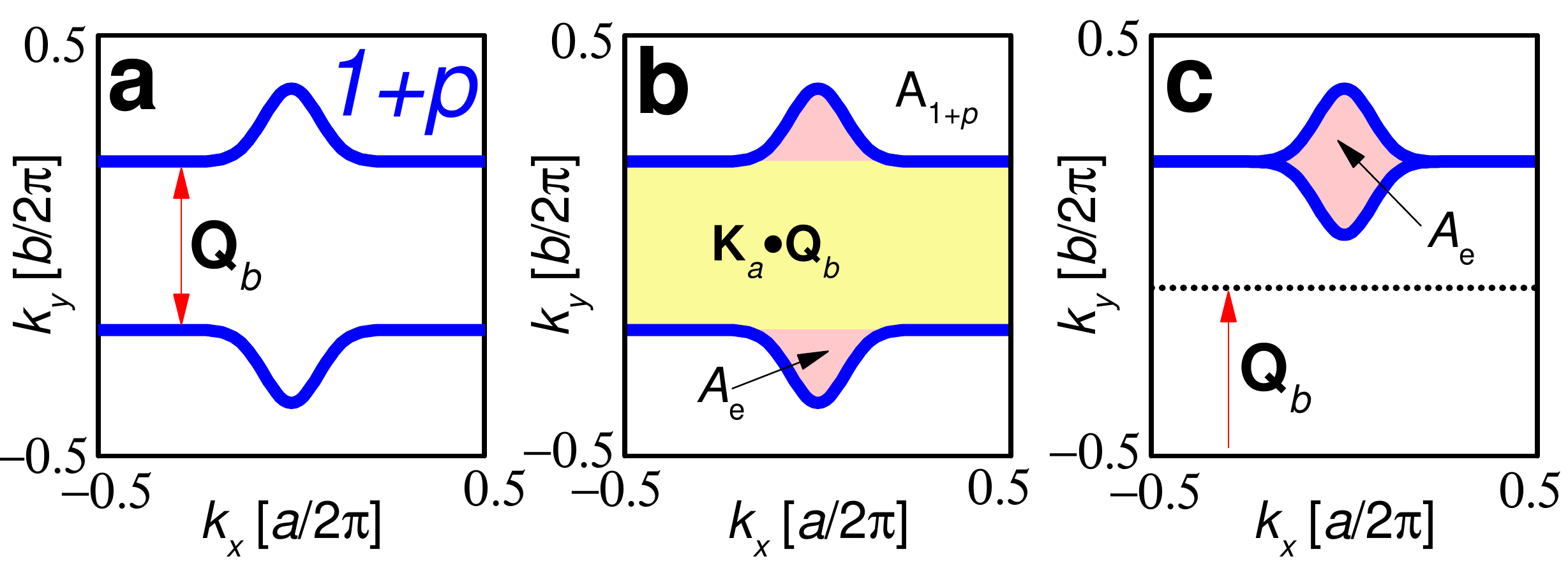}
\caption{({\bf a}), Schematic quasi-one-dimensional open Fermi surface (blue lines), with the charge-density wave ordering vector ${\bf Q}_b$ indicated (red). ({\bf b}), The same Fermi surface with the different area contributions shaded as described in the text. ({\bf c}), Schematic reconstructed electron pocket after translation of part of the unreconstructed Fermi surface.
}
\label{openFS}
\end{figure}

It is instructive to begin by considering the case of an open Fermi surface that becomes reconstructed by a unidirectional density-wave ordering vector (see Fig.~\ref{openFS}a). An imperfectly nested unreconstructed Fermi surface of equivalent topology occurs in quasi-one-dimensional organic conductors,\cite{chaikin1} and has also been proposed to occur in the cuprates when a large nematic distortion precedes the formation of a charge-density wave.\cite{yao1} We assume that the charge-density wave ordering vector ${\bf Q}_b=(0,\delta_b)\frac{2\pi}{b}$ spans the flat portions of the Fermi surface sheets in Fig.~\ref{openFS}a and that the carriers contained in the center of the Brillouin zone between the quasi-one-dimensional sheets are electrons. 

As shown in Fig.~\ref{openFS}b, the effect of the density-wave is to remove electrons from the unreconstructed Fermi surface that occupy an area equal to ${\bf K}_a\cdot{\bf Q}_b$ (indicated in yellow). After reconstruction, these electrons are accommodated within a series of completely filled reconstructed bands that lie below the chemical potential in the reconstructed electronic structure. On defining an irreducible rational fraction of the form $\delta_b=\frac{m_b}{n_b}$ for the ordering vector length, in which $m_b$ and $n_b$ are integers, a single band unreconstructed electronic structure is transformed into a reconstructed electronic structure consisting of $m_bn_b$ electronic bands. The electrons removed from the Fermi surface are then contained within $m_b$ completely filled bands that lie below the conduction band.

From visual inspection of Fig.~\ref{openFS}b, we see that the area ${\bf K}_a\cdot{\bf Q}_b$ (indicated in yellow), the total area $A_{\rm e}$ of the unnested portion of the electron Fermi surface (indicated in pink) and the area $A_{1+p}=\frac{1}{2}(1+p)A_{\rm UBZ}$ of the Brillouin zone occupied by holes (indicated in white) must together equal the area of the unreconstructed Brillouin zone $A_{\rm UBZ}={\bf K}_a\cdot{\bf K}_b$. Putting these terms together, we arrive at
\[{\bf K}_a\cdot{\bf Q}_b+A_{\rm e}+A_{1+p}=A_{\rm UBZ}\]
from which we obtain 
\begin{equation}\label{opendependence}
\delta_b^{\rm ~open}=\frac{1}{2}(1-p)-\frac{A_{\rm e}}{A_{\rm UBZ}}
\end{equation}
upon substituting ${\bf Q}_b$ and rearranging terms. The two unnested portions of the open Fermi surface in Fig.~\ref{openFS}b (indicated in pick) come together in Fig.~\ref{openFS}c to form a reconstructed Fermi surface consisting of a single electron pocket of area $A_{\rm e}$ (again, indicated in pink). The flat nested portions of the Fermi surface on either side of the reconstructed electron pocket in Fig.~\ref{openFS}c will disappear from the reconstructed Fermi surface upon the introduction of coupling terms linking the open sheets in the charge-density wave Hamiltonian. For $p=0$, the dependence of $\delta_b^{\rm ~open}$ on $A_{\rm e}$ is the same as that obtained in the quantized nesting model of magnetic field-induced-spin-density waves.\cite{chaikin1}

\subsection{Large Fermi surface and biaxial order}
\begin{figure}
\centering 
\includegraphics*[width=.45\textwidth]{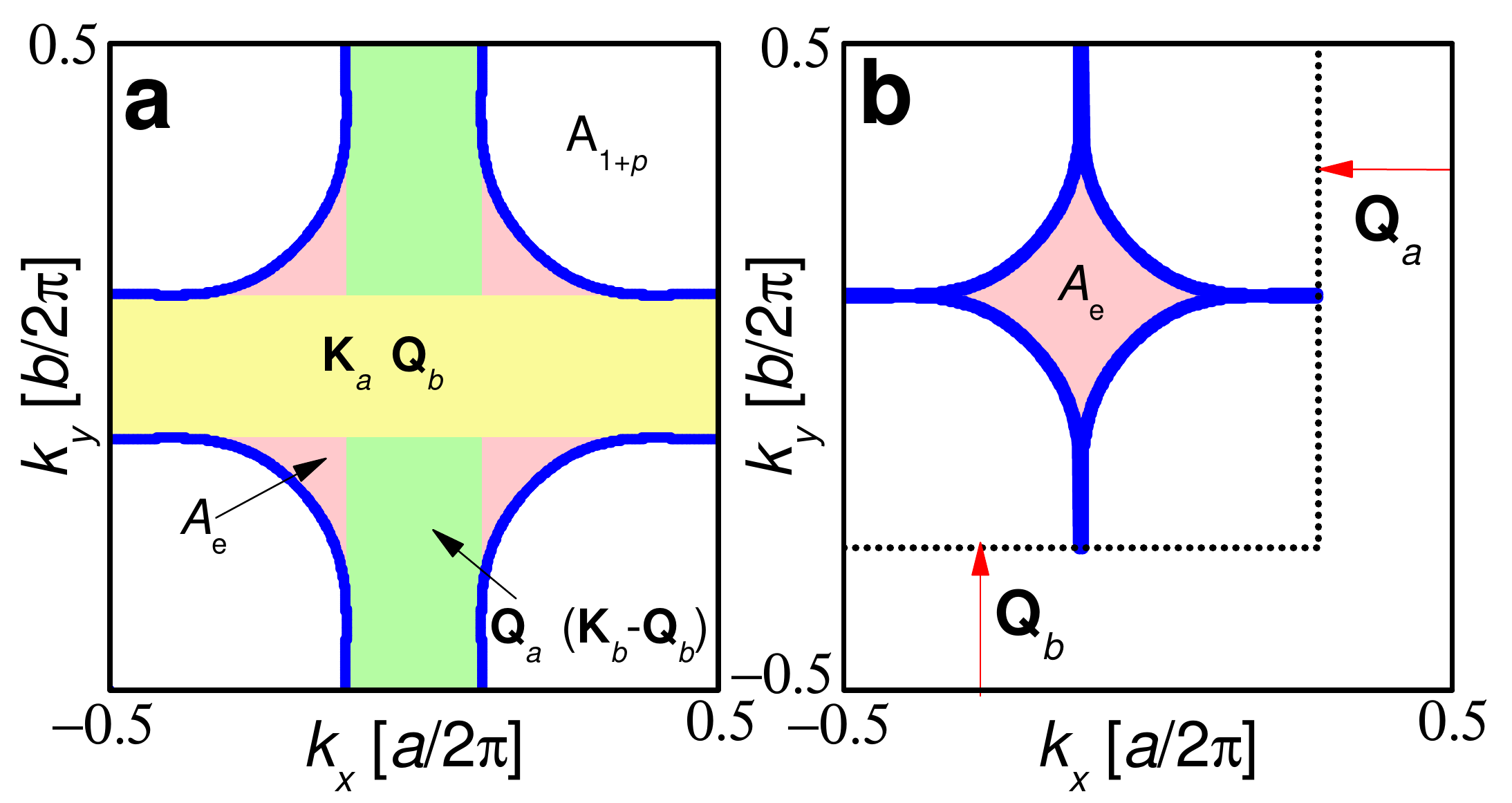}
\caption{({\bf a}), The large Fermi surface with the different area contributions shaded. ({\bf b}), Schematic showing the reconstructed electron pocket after translation of parts of the unreconstructed Fermi surface by ${\bf Q}_a$, ${\bf Q}_b$ and ${\bf Q}_a+{\bf Q}_b$.
}
\label{largeFS}
\end{figure}

On considering biaxial density-wave ordering starting from a large Fermi surface of the form shown in Fig.~\ref{startingFS}a, two ordering vectors ${\bf Q}_a$ and ${\bf Q}_b$ must now act in concert to remove electrons from the unreconstructed Fermi surface in Fig.~\ref{largeFS}a (indicated in yellow and green). We consider each of these in turn -- the precise order being unimportant. Starting with ${\bf Q}_b$, its effect is again to remove electrons from the unreconstructed Fermi surface occupying a total area area ${\bf K}_a\cdot{\bf Q}_b$ (indicated in yellow). The green and pink regions occupied by electrons survive this first step, but are folded by ${\bf Q}_b$ to produce multiple Fermi surfaces in higher order Brillouin zones (not shown for clarity). The effect of the second ordering vector ${\bf Q}_a$ is to remove remaining electrons from the unreconstructed Fermi surface occupying a total area ${\bf Q}_a({\bf K}_b-{\bf Q}_b)$ (indicated in green). Defining $\delta_a=\frac{m_a}{n_a}$ and $\delta_b=\frac{m_b}{n_b}$, the reconstructed electronic structure consists of a total of $m_am_bn_an_b$ reconstructed bands. The electrons removed from the unreconstructed Fermi surface will then be contained within $m_am_b(n_a+n_b-1)$ completely filled bands that lie below the conduction band of the reconstructed band structure.

Again, on equating all of these areas to $A_{\rm UBZ}$ in Fig.~\ref{largeFS}a we arrive at
\[{\bf K}_a\cdot{\bf Q}_b+{\bf Q}_a({\bf K}_b-{\bf Q}_b)+A_{\rm e}+A_{1+p}=A_{\rm UBZ}\]
from which we obtain
\begin{equation}\label{largedependence}
\delta^{\rm ~large}=1-\sqrt{\frac{1}{2}(1+p)+\frac{A_{\rm e}}{A_{\rm UBZ}}+d^2}
\end{equation}
on substituting ${\bf Q}_a$ and ${\bf Q}_b$ and rearranging terms. Here $\delta^{\rm ~large}$ refers to the average $\frac{1}{2}(\delta_a^{\rm ~large}+\delta_b^{\rm ~large})$ while $d$ refers to half the difference $\frac{1}{2}(\delta_a^{\rm ~large}-\delta_b^{\rm ~large})$. Since $d^2<$~10$^{-4}$, this term can mostly be neglected. The functional form of Equation (\ref{largedependence}) is identical to that obtained by way of a full numerical calculation in Ref.~\cite{harrison2}  -- where it was the period $\lambda=1/\delta^{\rm ~large}$ of the density-wave that was being plotted. The reconstructed Fermi surface in Fig.~\ref{largeFS}b has the same diamond-shaped electron pocket (indicated in pink) as discussed in several earlier biaxial reconstruction scenarios.\cite{harrison1,harrison2,sebastian1,sebastian2,maharaj1,allais1,harrison3,robinson1,briffa1} 

\subsection{Small Fermi surface and biaxial order}
\begin{figure}
\centering 
\includegraphics*[width=.45\textwidth]{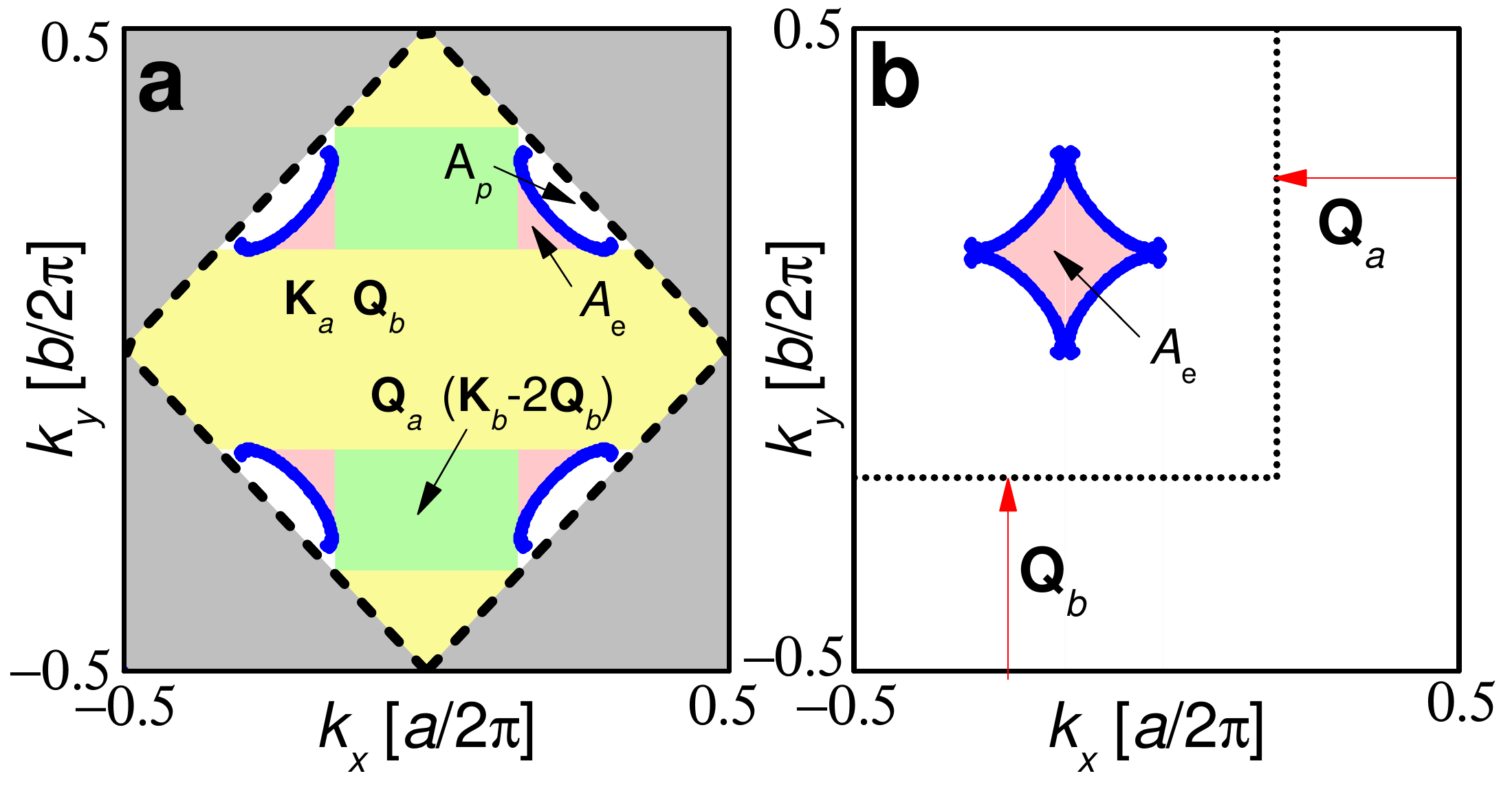}
\caption{({\bf a}), The small Fermi surface within the antiferromagnetic Brillouin zone with the different area contributions shaded. Grey indicates the regions outside the antiferromagnetic Brillouin zone. ({\bf b}), Schematic showing the reconstructed electron pocket after translation of parts of the unreconstructed Fermi surface by ${\bf Q}_a$, ${\bf Q}_b$ and ${\bf Q}_a+{\bf Q}_b$.
}
\label{smallFS}
\end{figure}

On considering biaxial density-wave ordering starting from a small Fermi surface of the form shown in Fig.~\ref{startingFS}b, we proceed to sum the areas in the same way as we would for a density-wave coexisting with ${\bf Q}_{\rm AFM}=(\pi,\pi)$ antiferromagnetic order. In this case we sum the areas within the antiferromagnetic Brillouin zone of area $A_{\rm ABZ}=\frac{1}{2}A_{\rm UBZ}$ (see Fig.~\ref{smallFS}a) and consider only the lower of the two bands in the antiferromagnetic starting electronic structure. We must therefore also neglect the regions outside antiferromagnetic Brillouin zone shaded in grey. The area $A_p=\frac{p}{2}A_{\rm UBZ}$ of the Brillouin zone occupied by holes (indicated in white) is now significantly smaller than before. 

The effect of ${\bf Q}_b$ is once again to remove electrons from the unreconstructed Fermi surface occupying a total area area ${\bf K}_a\cdot{\bf Q}_b$ (indicated in yellow). This time, the second vector ${\bf Q}_a$ removes electrons occupying a remaining area of ${\bf Q}_a({\bf K}_b-2{\bf Q}_b)$ (indicated in green). On equating all of these areas to $\frac{1}{2}A_{\rm ABZ}$ in Fig.~\ref{largeFS}b we arrive at
\[{\bf K}_a\cdot{\bf Q}_b+{\bf Q}_a({\bf K}_b-2{\bf Q}_b)+A_{\rm e}+A_{p}=\frac{1}{2}A_{\rm UBZ}\]
from which we obtain
\begin{equation}\label{smalldependence}
\delta^{\rm ~small}=\frac{1}{2}-\sqrt{\frac{1}{2}\bigg(\frac{p}{2}+\frac{A_{\rm e}}{A_{\rm UBZ}}\bigg)+d^2}
\end{equation}
on substituting ${\bf Q}_a$ and ${\bf Q}_b$ and rearranging terms. Here, similar to what we have for the large Fermi surface, $\delta^{\rm ~small}=\frac{1}{2}(\delta_a^{\rm ~small}+\delta_b^{\rm ~small})$ while $d=\frac{1}{2}(\delta_a^{\rm ~small}-\delta_b^{\rm ~small})$. The reconstructed Fermi surface in Fig.~\ref{smallFS}b continues to have the same diamond-shaped electron pocket (indicated in pink) as discussed in several earlier publications.\cite{harrison1,harrison2,sebastian1,sebastian2,maharaj1,allais1,harrison3,robinson1,briffa1}

\subsection{Generalized Fermi surface considerations}
In the case of more generalized forms for the large unreconstructed hole Fermi surface in the cuprates, the outcome will depend on the strength of the charge-density wave coupling. In the weak coupling limit, imperfect nesting produces additional small sections of Fermi surface. Examples of such pockets are described for the case of biaxial charge-density wave ordering in Refs.~\cite{harrison1,allais1} The areas of these sections of Fermi surface must be respectively added to or subtracted from $A_{\rm e}$, depending or whether they contain electrons or holes. Hole pockets like those discussed in Ref.\cite{allais1} cannot contribute to $\delta^{\rm~small}$, however, as these would lie mostly outside of the antiferromagnetic Brillouin zone.

Small additional sections of Fermi surface are less likely to occur as the strength of the charge-density wave coupling is increased. As the coupling is progressively increased, the Fermi surface will eventually consist of a single reconstructed electron pocket.\cite{harrison1} At this point one can then draw shaded regions with areas equal to those in Figs.~\ref{openFS}b and \ref{largeFS}a that add up to $A_{\rm UBZ}$. The shapes are nevertheless likely to become more irregular. There are two reasons why Equations (\ref{opendependence}) and (\ref{largedependence}) continue to be valid in the strong coupling limit. The first is that the chemical potential always adjusts itself to maintain the area $A_{1+p}$ of the Brillouin zone occupied by holes at a value compatible with the hole doping. The area $A_{1+p}$ is therefore invariant under an increase in the coupling strength. The second reason is that each cycle of a spin- or charge-density wave state always contains precisely an even number of electrons or holes and increments the phase by $2\pi$. The number of electrons removed from the Fermi surface by a density-wave state, and the area that they occupy within the Brillouin zone, is therefore also independent of the strength of coupling. The pocket area $A_{\rm e}$, meanwhile, is constrained by Onsager's relation $A_{\rm e}=2\pi eF_{\rm e}/\hbar$, where $F_{\rm e}$ is the measured quantum oscillation frequency.

Bilayer coupling will have different effects on the doping dependence of $\delta$, depending on its strength compared to the strength of the spin- or charge-density wave coupling, or depending on whether the density-wave connects bands of the same or opposite parity.  If the bilayer coupling is much weaker than the density-wave coupling, or if the density-wave connects bilayer-split bands of opposite parity,\cite{briffa1} then there will continue to be single values of $\delta^{\rm~large}$ and $\delta^{\rm~small}$. Briffa~{\it et al.}\cite{briffa1} have shown that in the case where the density-wave connects bands of opposite parity, two degenerate reconstructed Fermi surfaces are obtained that are related to each other by way of a 90$^\circ$ rotation. If, on the other hand, bilayer coupling is large and the density-wave connects bilayer-split bands of like parity,\cite{harrison4} one will then find that $\delta$, $A_{\rm e}$ and $A_{1+p}$ can each have different values for the bonding and antibonding bands. Recent x-ray scattering studies indicate a broken mirror plane orthogonal to the $c$ axis centered on the bilayer in YBa$_2$Cu$_3$O$_{6+x}$,\cite{forgan1,chang2} which supports a scenario in which density-wave connects bands of opposite parity.\cite{briffa1}

\section{Comparison with experiment}
\begin{figure}
\centering 
\includegraphics*[width=.43\textwidth]{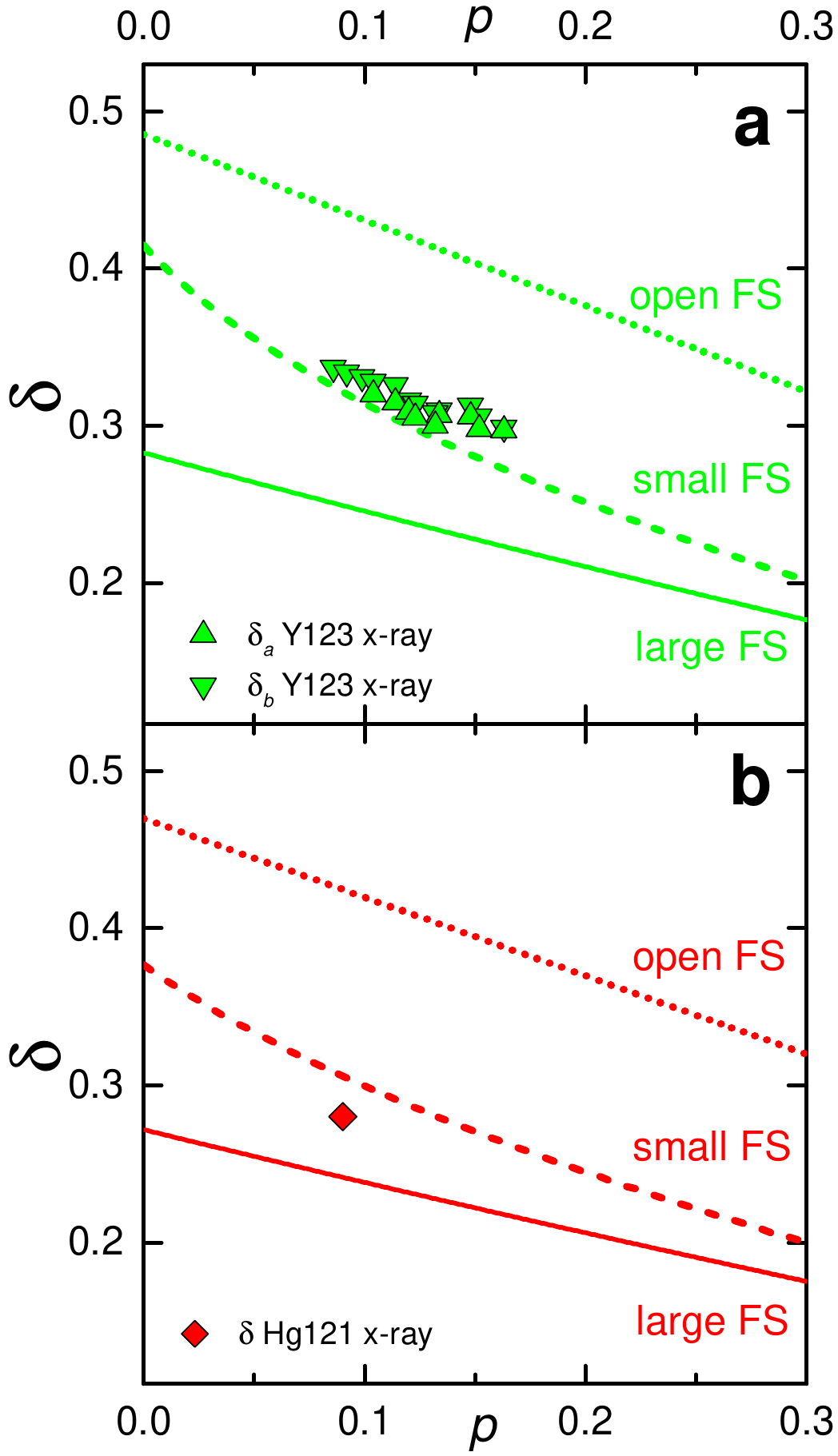}
\caption{({\bf a}), A comparison of $\delta$ calculated according to the three different open (dotted line), large (solid line) and small (dashed line) Fermi surface models [using Equations (\ref{opendependence}), (\ref{largedependence}) and (\ref{smalldependence})] with experimental $\delta_{a,b}$ values obtained using x-ray scattering, as indicated. Data are shown for YBa$_2$Cu$_3$O$_{6+x}$ (Y123)\cite{blanco1,blackburn1} in ({\bf a}) and HgBa$_2$CuO$_{4+x}$ (Hg121)\cite{tabis1} in ({\bf b}).}
\label{delta}
\end{figure}

Figure~\ref{delta} shows the doping-dependence of $\delta_b^{\rm open}$, $\delta^{\rm large}$ and $\delta^{\rm small}$ calculated using Equations (\ref{opendependence}), (\ref{largedependence}) and (\ref{smalldependence}), respectively, neglecting $d^2$. In Fig.~\ref{delta}a we compare the lengths of the charge-density wave ordering vectors $\delta_a$ and $\delta_b$ obtained in x-ray scattering experiments\cite{blanco1,blackburn1} in YBa$_2$Cu$_3$O$_{6+x}$ with those calculated using the ratio $\frac{A_{\rm e}}{A_{\rm UBZ}}$ obtained from magnetic quantum oscillation experiments (assuming Onsager's relation).\cite{ramshaw2}  We assume a single pocket per CuO$_2$ plane\cite{briffa1} and approximate the doping-dependent quantum oscillation frequency in Ref.\cite{ramshaw2} with a linear fit, from which we obtain $F_{\rm e}\approx(399+1288~p)$~T. In Fig.~\ref{delta}b we compare the length of the charge-density wave ordering vector $\delta_a$ obtained in an x-ray scattering experiment\cite{tabis1} in HgBa$_2$CuO$_{4+x}$, assuming that $\delta_b=\delta_b$ in the tetragonal crystal structure, with those calculated using the ratio $\frac{A_{\rm e}}{A_{\rm UBZ}}$ obtained from magnetic quantum oscillation experiments.\cite{barisic1} In this case $F_{\rm e}\approx$~840~T is the quantum oscillation frequency measured at a single value of the hole doping. 

\section{Discussion}
It is clear from Fig.~\ref{delta} that the experimentally observed values of the charge-density wave ordering vector lengths $\delta_a$ (and $\delta_b$ in the case of YBa$_2$Cu$_3$O$_{6+x}$), are much shorter than those $\delta^{\rm~open}$ expected for a nematically-deformed Fermi surface accompanied by the formation of a unidirectional charge-density wave producing a single reconstructed Fermi surface pocket. The observed values of the charge-density wave ordering vector lengths are also found to be much longer than those $\delta^{\rm~large}$ expected for biaxial order producing a single reconstructed Fermi surface pocket starting from a large unreconstructed Fermi surface comprising $1+p$ hole carriers, as predicted by band structure calculations. Only by considering a small starting Fermi surface comprising $p$ carriers, do we find the predicted ordering vector lengths $\delta^{\rm~small}$ to be consistent with $\delta_{a,b}$ both at a quantitative level and in the overall form of its doping dependence. 

Some degree of discrepancy between $\delta_{a,b}$ and $\delta^{\rm~small}$ could potentially originate from assumptions that are made to evaluate the chemical hole doping, or to the presence of additional, as yet unobserved, small Fermi surface pockets. In YBa$_2$Cu$_3$O$_{6+x}$, for example, the hole doping is estimated by comparing the doping dependence of the rescaled superconducting transition temperature $T_{\rm c}$ with that of La$_x$Sr$_{2-x}$CuO$_4$.\cite{liang1} In HgBa$_2$CuO$_{4+x}$, meanwhile, the maximum quantum oscillation amplitude and plateaux in $T_{\rm c}$ versus $p$ are found to occur near $p\approx$~0.09\cite{chan1} rather than $p\approx$~0.12 in YBa$_2$Cu$_3$O$_{6+x}$. Closer agreement with $\delta^{\rm~small}$ would be obtained in Fig.~\ref{delta}b were $p$ estimated using the same method as used for YBa$_2$Cu$_3$O$_{6+x}$.\cite{liang1} %a hole doping closer to $p=0.12$ would apply in Fig.~\ref{delta}b, which would then bring this data point into closer agreement with $\delta^{\rm~small}$.

\begin{figure}
\centering 
\includegraphics*[width=.43\textwidth]{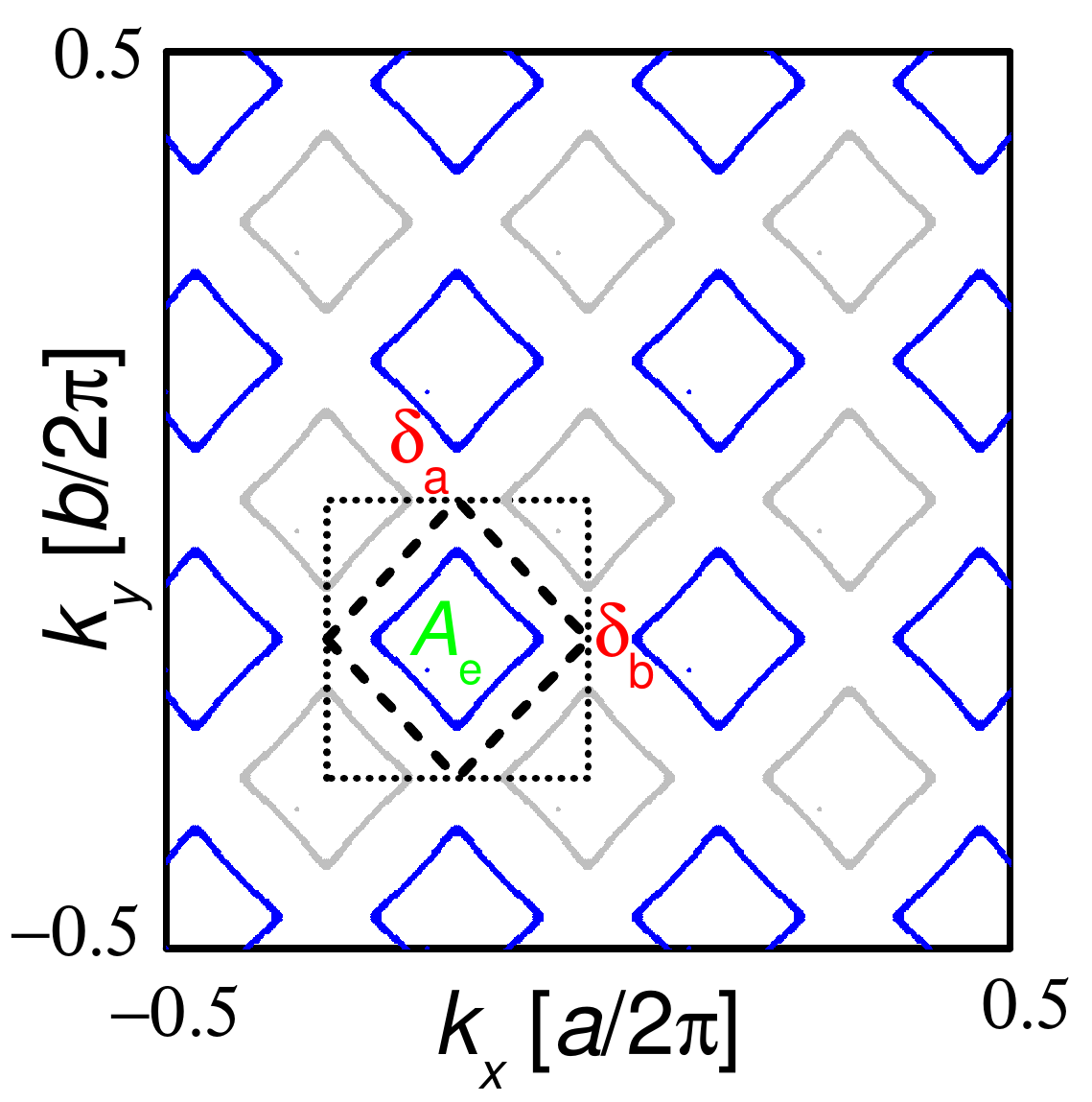}
\caption{Schematic reconstructed Fermi surface in the repeated Brillouin zone representation, with the electron pocket indicated in blue and the folded antiferromagnetic Brillouin zone boundary indicated by a dashed line. In all reconstruction scenarios involving the small starting Fermi surface, the states along the folded antiferromagnetic Brillouin zone become gapped by the charge-density wave order. If the wave vector  ${\bf Q}_{\rm AFM}$, in addition to ${\bf Q}_a$ and ${\bf Q}_b$, is involved in reconstructing the Fermi surface, then the dashed line becomes the true reconstructed Brillouin zone boundary and additional instances of the reconstructed electron pocket will appear (depicted in grey).}
\label{pocket}
\end{figure}

The form of the electronic dispersion at the antiferromagnetic Brillouin zone boundary is unique for each model of the small starting Fermi surface.\cite{qi1,lee1,chakravarty1,rice1,chowdhury1} However, since it is the states close to the antiferromagnetic Brillouin zone boundary that become gapped by the charge-density wave (see Fig.~\ref{pocket}), the unique differences in their dispersions are essentially lost once Fermi surface reconstruction takes place. The primary role of the Coulomb repulsion in all of these models is therefore only to provide a mechanism for the opening of a large gap in the in the antinodal region of the Brillouin zone.  

The close correspondence of $\delta_{a,b}$ with $\delta^{\rm small}$ expected for a small Fermi surface implies that the experimental values of $\delta_{a,b}$ and $A_{\rm e}$ can be used to to obtain the number of holes contained within the Fermi surface. On rearranging 
the terms in Equation~(\ref{smalldependence}) and using $n_{\rm h}=2A_p/A_{\rm UBZ}$, we obtain
\begin{equation}\label{inverse}
n_{\rm h}=4\bigg[\bigg(\frac{1}{2}-\delta\bigg)^2-d^2-\frac{A_{\rm e}}{2A_{\rm UBZ}}\bigg].
\end{equation}
In Fig.~\ref{startingFS}c, we compare the experimental estimates of $n_{\rm h}$ against those $n_{\rm h}=p$ and $n_{\rm h}=1+p$ expected for the small and large Fermi surface, respectively. A continuation of the linear trend $n_{\rm h}=p$ previously obtained for very low hole dopings $p<$~0.08\cite{segawa1} is suggested, followed by a sharp jump by one hole per CuO$_2$ plane near optimal doping to arrive at $n_{\rm h}=1+p$. 

Measurements of the Hall coefficient $R_{\rm H}$ have suggested an increase in $n_{\rm h}$ near optimal dopings,\cite{badoux1} although they have also suggested the crossover between $n_{\rm h}=p$ and $n_{\rm h}=1+p$ to occur over an extended range of dopings 0.15~$<p<$~0.20.  Several factors, including changes curvature around the Fermi surface and anisotropic scattering rates,\cite{caprara1,harrison3,lin1,hussey1} cause 
$R_{\rm H}$ no longer to be directly related to the number of carriers or the sign of the carriers contained within the Fermi surface once $\omega_{\rm c}\tau\lesssim$~1, where $\omega_{\rm c}$ is the cyclotron frequency and $\tau$ is the scattering time. This situation is more likely to apply near optimal doping owing to the increase in $\tau^{-1}$. %Fermi surface curvature, magnetic breakdown tunneling and fluctuations are known to cause the Hall coefficient to depart from the simple Drude form expected for a circular Fermi surface.\cite{caprara1,harrison3,lin1}
%While a small starting Fermi surface appears to be indicated, we cannot identify a specific model.

\section{Conclusion}
Having considered a large starting Fermi surface comprising $n_{\rm h}=1+p$ hole carriers, as predicted by band structure calculations, and a small starting Fermi surface comprising $n_{\rm h}=p$ carriers, as expected in the presence of antiferromagnetic correlations, we find the small starting Fermi surface to show a close correspondence with the lengths of the wave vectors $\delta_{a,b}$ obtained from x-ray diffraction experiments within the underdoped regime over a broad range of hole dopings. The reconstructed Fermi surface seen in magnetic quantum oscillation and the `Fermi arcs' seen in angle-resolved photoemission spectroscopy\cite{hossain1} measurements must therefore originate from the same small starting Fermi surface. Our findings imply that quantum oscillation frequency and charge-density wave vectors can be used to provide an accurate means for estimating the number of holes contained within the Fermi surface over the majority of the pseudogap regime in the low temperature limit.

A small starting Fermi surface consisting of four hole pockets (e.g. Fig.~\ref{startingFS}b) is expected to be one of the consequences of the on-site Coulomb repulsion continuing to remain dominant over low energy excitations when holes are doped into the Mott insulator.\cite{chakravarty1,lee1,qi1,rice1,chowdhury1} Our identification of the small Fermi surface coexisting with charge-density wave order implies that the Coulomb repulsion must remain the dominant energy scale throughout the entire underdoped regime. It must therefore also play an essential role in the formation of the pseudogap and in driving significant changes in the Fermi surface and quantum critical behavior close to optimal doping.\cite{ramshaw2}
\\

\section{Acknowledgements}
This work is supported by %the Royal Society, King's College (Cambridge University), 
the US Department of Energy BES ``Science at 100 T" grant no. LANLF100, the National Science Foundation and the State of Florida. %SES acknowledges funding from the Royal Society and
%the European Research Council under the European Union's Seventh
%Framework Programme (FP/2007-2013) / ERC Grant Agreement no.
%337425-SUPERCONDUCTINGMOTT.

\end{document}